\documentclass[acmsmall,screen]{acmart}
\AtBeginDocument{%
  \providecommand\BibTeX{{%
    \normalfont B\kern-0.5em{\scshape i\kern-0.25em b}\kern-0.8em\TeX}}}

\acmConference[CSCW '24]{San José '24: ACM Conference On Computer-Supported Cooperative Work And Social Computing}{November 9--13, 2024}{San José, Costa Rica}
\acmPrice{15.00}
\acmISBN{978-1-4503-XXXX-X/18/06}

\acmSubmissionID{2343}

\usepackage{fontawesome,dirtytalk}
\usepackage{xcolor}
\usepackage{graphicx,multirow,array}
\usepackage{color,soul}
\usepackage{tikz}
\usepackage{libertine}
\usepackage{rotating,tabularx}

\begin{document}


\title[A Case Study of Open-Source
Hardware Design Collaboration]{``A Lot of Moving Parts'': A Case Study of Open-Source Hardware Design Collaboration in the Thingiverse Community}



\author{Kathy Cheng}
\affiliation{%
  \institution{University of Toronto}
  \streetaddress{27 King's College Circle}
  \city{Toronto}
  \country{Canada}}
\email{kathy.cheng@mail.utoronto.ca}

\author{Shurui Zhou}
\affiliation{%
  \institution{University of Toronto}
  \city{Toronto}
  \country{Canada}}

\author{Alison Olechowski}
\affiliation{%
  \institution{University of Toronto}
  \city{Toronto}
  \country{Canada}}

\renewcommand{\shortauthors}{Kathy Cheng et al.}

\begin{abstract}
Open-source is a decentralized and collaborative method of development that encourages open contribution from an extensive and undefined network of individuals. Although commonly associated with software development (OSS), the open-source model extends to hardware development, forming the basis of open-source hardware development (OSH). Compared to OSS, OSH is relatively nascent, lacking adequate tooling support from existing platforms and best practices for efficient collaboration. Taking a necessary step towards improving OSH collaboration, we conduct a detailed case study of \textit{DrawBot}, a successful OSH project that remarkably fostered a long-term collaboration on Thingiverse -- a platform not explicitly intended for complex collaborative design. Through analyzing comment threads and design changes over the course of the project, we found how collaboration occurred, the challenges faced, and how the DrawBot community managed to overcome these obstacles. Beyond offering a detailed account of collaboration practices and challenges, our work contributes best practices, design implications, and practical implications for OSH project maintainers, platform builders, and researchers, respectively. With these insights and our publicly available dataset~\cite{cheng_data_2024}, we aim to foster more effective and efficient collaborative design in OSH projects.

\end{abstract}



\maketitle

\section{Introduction} 

From smartphones to automobiles, coffee makers to pacemakers, physical products are indispensable in our daily lives. Designing and developing these products successfully requires effective collaboration among multiple individuals~\cite{bellotti_walking_1996,Pawlak2010}. Complex hardware development has been traditionally restricted to professional engineering firms, performed by design teams operating within formal organizational structures and commercial incentives~\cite{Ulrich2015}. However, the open-source model has recently emerged as a promising alternative means of hardware development.

Open-source is a decentralized and collaborative method of development that encourages open contribution from an extensive and undefined network of individuals~\cite{Olson2012,germonprez_eight_2018}. Contributors are distributed physically in time and space, membership is transient (which allows contributors to come and go as they please), and agents are semi-autonomous in their work~\cite{gero_collective_2015}. Open-source communities have the freedom to follow non-profit motivations~\cite{Bonaccorsi2003}, which may lead to design innovations for under-served or low-power populations. 

The open-source model is often associated with software development (OSS), and this
success~\cite{Osterloh2007} has inspired product designers to seek a similar approach to hardware development (i.e., OSH)~\cite{herrera_promises_2020}. Although the goals of open-source are shared among hardware and software, the artifacts involved in the design processes are distinct; while OSS mainly relies on text-based source code, OSH projects involve 3D CAD (computer-aided design) models that represent geometric and topological data~\cite{stirling_hardops_2022,cheng_age_2023}. Due to differences in the artifacts, OSH requires considerations for design representation, materials, and manufacturing~\cite{antoniou_identifying_2021}. Therefore, while lessons on collaboration effectiveness and efficiency from OSS are relevant to OSH, the tools and practices cannot be directly applied. 



Many hardware designers are enthusiastic about OSH~\cite{bonvoisin_how_2019}; however, the potential for complex collaborative design is currently limited by the lack of widely-accepted, dedicated groupware for OSH~\cite{bonvoisin_current_2017,mies_development_2020}. As a result, some hardware designers have resorted to using social coding platforms like GitHub~\cite{bonvoisin_how_2019,dai_correlation_2020}, despite not being intended for hardware artifacts like 3D CAD models.
Predictably, a software-focused platform is ill-suited for hardware development and introduces pain points and collaboration inefficiencies, such as cumbersome 3D model manipulation~\cite{bonvoisin_current_2017}, poor granularity for branch/merge~\cite{cheng_user_2023}, and lack of synchronous/asynchronous communication channels~\cite{mies_development_2020}. As OSH expands, it is imperative to devise dedicated strategies to facilitate collaboration within OSH projects, rather than inheriting existing systems from OSS.

Platforms dedicated to hardware do exist, in the form of \textit{collective design} communities (e.g., Thingiverse,\footnote{https://thingiverse.com/} Thangs3D\footnote{https://thangs.com/}), as opposed to  \textit{collaborative design} communities. The primary aim of these platforms is for sharing designs for 3D printed artifacts, rather than supporting a complex collaborative design process~\cite{dai_issues_2020}. However, anecdotal evidence suggests that makers are indeed trying to engage in collaborative design activities through reusing existing designs -- stretching the collective design platform's capabilities to enable coordinated action -- despite the limited collaboration features (e.g., poor version tracking, lack of multi-user CAD)~\cite{oehlberg_patterns_2015,reddit2023,li_dynamic_2023}. 

In light of the growing interest in collaborative OSH and anecdotes of successful projects in spite of the insufficient support from existing platforms, there is a crucial need to comprehend how collaboration unfolds in OSH projects. Current empirical research on OSH has focused on general patterns of design in collective design communities, and pain points faced by individual makers~\cite{alcock_barriers_2016,hudson_understanding_2016}. Although these studies are useful for characterizing high-level practices, they lack in-depth exploration of the dynamics, processes, or inefficiencies specific to collaboration. Without this understanding, it remains unclear how to develop best practices or dedicated tooling support for efficient and effective OSH collaboration.

Our goal is to close this knowledge gap so that we can better understand and improve the way that makers collaborate on OSH designs. We first need to provide a descriptive analysis of the current practices that enable successful collaborative design, and then we can identify the challenges faced in the community. The following two research questions guided our work:

\begin{itemize}
  \item[\textbf{RQ1:}] How do designers collaborate during a successful open-source hardware (OSH) project?
  \begin{itemize}
      \item[RQ1.1:] What tasks involved collaboration?
      \item[RQ1.2:] How did designers collaborate to achieve those tasks?
  \end{itemize}
  \item[\textbf{RQ2:}] How is OSH collaboration hindered, and what are the opportunities for improvement?
\end{itemize}

To address these questions, we conducted a detailed case study of \textit{``DrawBot''}, an OSH project hosted on Thingiverse, the largest online 3D printing community~\cite{thingiversecom_thingiverse_2023}. Although Thingiverse is primarily a collective design platform -- and not explicitly intended for complex OSH collaboration -- the case we present serves as a rare instance of thriving collaborative design. The case study approach enables a retrospective analysis of a collaborative, long-lived, and well-maintained OSH project, unveiling the aspects contributing to the project's success, and areas for improvement in future projects. Focusing on a project-level analysis allows for a deeper investigation of the collaborative interactions among participants, providing longitudinal insights into how both the design and community evolved over the project's lifetime. While previous research has applied the case study method to examine open-source collaboration in software communities~\cite{li_how_2021,Mockus2000,Gharehyazie2014}, the exploration of OSH collaboration practices is relatively nascent. To our knowledge, our study is the first to investigate collaborative design within a hardware-focused community. Therefore, we make the following contributions to CSCW:

\begin{enumerate}
    \item An in-depth case study of a unique collaborative OSH design project that successfully overcame tooling limitations to flourish on a collective design platform.
    \item An investigation of how makers collaborated on both design-related and non-design-related tasks, offering best practices for the maintainers and participants of OSH projects to collaborate efficiently. 
    \item An identification of the tooling support limitations which harm collaborative awareness and impede the full potential of OSH collaboration. As a first step towards addressing these challenges, our work contributes several design implications for OSH platform builders.
    \item A novel approach for analyzing collaborative design projects on collective design platforms. \textcolor{black}{To facilitate further exploration of OSH collaboration, we provide practical insights for researchers conducting similar case studies, as well as our publicly available dataset (found here~\cite{cheng_data_2024}).}
\end{enumerate}

\section{Background \& Related Work}\label{sec_background}

\subsection{Open Source Hardware Development (OSH)}\label{OSHW}
Open-source hardware development (OSH) allows free and open participation from a distributed, asynchronous, and large network of users~\cite{ozkil_collective_2017}. Products developed through OSH efforts are physical artifacts that are free for the public to: (1) \textit{study} (i.e., access sufficient information to understand a product's use and design), (2) \textit{modify} (i.e., create product derivatives), (3) \textit{make} (i.e., physically fabricate and manufacture the product), and (4) \textit{distribute} (i.e., share or sell the physical product or associated documentation)~\cite{OSHW2023}. 

There is no single, widely-accepted platform for OSH, comparable to GitHub's widespread use in OSS~\cite{escamilla_rise_2022}. OSH occurs on various platforms, such as Thingiverse~\cite{oehlberg_patterns_2015,berman_thingipano_2020,buehler_sharing_2015}, GitHub~\cite{bonvoisin_how_2019,dai_correlation_2020}, and Instructables~\cite{tseng_product_2014}. OSH projects typically leverage an assortment of tools for collaboration, communication, or version control~\cite{bonvoisin_what_2017}. Mies et al. evaluated five popular platforms used for OSH based on their support of communication, coordination, knowledge integration, and information logistics~\cite{mies_development_2020}.\footnote{The five platforms analyzed by Mies et al. were GitHub, MediaWiki, GrabCAD, Wevolver, and Wikifactory.} GitHub narrowly outperformed the other OSH-specific options, despite not being designed to support collaboration on 3D CAD models. In fact, none of the five platforms offered collaborative modelling capabilities~\cite{mies_development_2020}. 

\subsubsection{Collective Design vs. Collaborative Design}
OSH practices have been classified in the literature as following a variety of working styles, such as, open design~\cite{bakirlioglu_framing_2019,tooze_open_2014}, collective design~\cite{paulini_design_2013,ozkil_collective_2017,gero_collective_2015,hofmann_making_2022}, peer-to-peer design~\cite{bonvoisin_how_2019}, social product design~\cite{abhari_co-innovation_2017}, and collaborative design~\cite{stirling_hardops_2022,dai_correlation_2020}. While there is ambiguity in the terminology of characterizing types of OSH projects, scholars generally agree that OSH projects can be broadly categorized into two non-mutually exclusive types, which we call \textit{collective design} and \textit{collaborative design}. 

Collective design occurs when large numbers of people engage in design activity to accomplish their individual goals~\cite{gero_collective_2015}. Designers may only know each other through their designs, and communicate through loosely formed and informal design networks~\cite{gero_collective_2015}. On collective design platforms (e.g., Thingiverse, Thangs3D), users can freely post designs, contribute to other users' designs, and build upon each others' designs in a sequential and derivative manner~\cite{ozkil_collective_2017}. While collaboration occurs in the sense that the design process involves multiple people, these processes do not involve coordinated action that evolves the design towards a shared goal~\cite{bonvoisin_how_2019}.

In contrast, collaborative design involves the joint efforts of many designers to address design goals, requiring coordination and cooperation~\cite{paulini_design_2013,tooze_open_2014}. Products designed in this manner are typically complex -- in both product architecture and required effort~\cite{jacobs_product_2007}, mimicking products developed in industrial product design teams~\cite{bonvoisin_what_2017}, thus necessitating collaborative design. Unlike in formal design organizations, participants of open-source collaborative design are distributed in time and space, self-organized, and transient in their level and duration of involvement. Nonetheless, a common motivation, communication channels, and shared representation (i.e., design artifacts) facilitate the ongoing collaborative design process~\cite{maher_scaling_2011}. 

Previous research has investigated the evolutionary design patterns within collective design platforms (as we will review in Section~\ref{thingiverse}), however, there are few empirical accounts of collaborative design in OSH. Müller-Seitz and Reger documented a case study of the OScar project, an open-source effort to design and develop a functional car through the Internet~\cite{muller-seitz_networking_2010}. The study demonstrated that tooling support, in the form of discussion boards, was of utmost importance to the coordination of the project. However, both the project and the investigation predated many ICT (information and communication technology) and CMC (computer-mediated communication) tools that are available today, such as wiki software, video-sharing platforms, or blogging platforms~\cite{muller-seitz_networking_2010}. An updated case study is needed to describe current OSH collaboration practices and challenges.

More recently, Mellis and Buechley explored OSH collaboration on the Arduino microcontroller development platform, making comparisons to OSS~\cite{mellis_collaboration_2012}. The Arduino community resembled an ecosystem of many small-scale collaborations among groups of contributors. This decentralized nature was reflected in the design process, where contributors created many design alternatives, rather than improving one single, central option. Making alternative designs was considered a positive contribution to the community, in contrast to OSS, where divergent repositories, or forks, are often seen as resulting from disagreement or conflict among developers~\cite{mellis_collaboration_2012,WurzelGonalves2022}, which can fragment the community~\cite{zhou_what_2019}.

\subsubsection{Collaborative Design Challenges in OSH}

There is a desire for collaborative OSH, but there is not yet a suitable platform to seamlessly accommodate collaborative design. Some past studies have pointed out the challenges of OSH, which we summarize below. 

OSH projects inherently involve diverse artifacts, encompassing both physical and digital components, which distinguish them from OSS projects. While OSS users can compile code at virtually zero cost to produce the final product, OSH users face challenges related to the cost of materials and manufacturing for physical components~\cite{howard_open_2012,antoniou_identifying_2021,hansen_current_2013}. These considerations not only apply to the end product, but also at various prototyping stages, which may discourage makers from continuously testing new design iterations~\cite{bonvoisin_how_2019}. Furthermore, the physical nature of hardware products prevents distributed collaborators from being able to test each other's design changes~\cite{mellis_collaboration_2012}.

Collaboration is also challenging with the digital artifacts in OSH projects, such as 3D CAD models. The lack of in-browser integration of 3D models is a frequently cited barrier to OSH collaboration~\cite{bonvoisin_current_2017}. Current platforms offer little opportunity for viewing and manipulating CAD models (e.g., rotation, annotation, exploded view), interoperability (compatibility with various 3D file formats), and online concurrent editing~\cite{bonvoisin_current_2017}.

Beyond artifact-related challenges, OSH faces many process-related challenges that hinder collaborative awareness~\cite{dourish_awareness_1992}. One major challenge is the lack of version control systems dedicated to OSH, resulting in the adoption of tools intended for software development. These tools simply store subsequent versions of a file without offering mechanisms to summarize changes between versions~\cite{mellis_collaboration_2012}. The inadequate tracking of design history versions makes it difficult to reuse models~\cite{ezoji_towards_2021} -- which is particularly harmful to OSH projects, where the creation of derivatives from existing designs is a fundamental practice~\cite{mellis_collaboration_2012,ozkil_collective_2017}.

Documentation is of utmost importance to open-source collaboration, allowing for a simple way to make knowledge accessible to a large community~\cite{hansen_current_2013}. Compared to OSS, OSH employs a diverse range of documentation types, encompassing images, CAD files, and videos~\cite{mack_rapid_2023}. This diversity adds complexity to collaboration since processes with physical artifacts (e.g., assembly) are not inherently captured in a digital format~\cite{tseng_product_2014}.

Finally, the absence of one universally accepted OSH platform results in collaboration occurring simultaneously on different platforms, such as social media, forums, blogs, GitHub, or Thingiverse~\cite{dai_correlation_2020}. With the scattered nature of the project, OSH participants must navigate all of these different sites, which can harm collaboration efficiency.

Overall, existing research has characterized OSH practices, offering two conceptions: collective design and collaborative design. OSH has primarily fostered collective design, where individuals are accomplishing their own goals on the platform, whereas less attention has been given to collaborative design, which faces many coordination and cooperation challenges. To address this problem, our case study contributes a detailed investigation of a collaborative design project, which, despite being conducted on a collective design platform, has successfully fostered a long-term, complex collaboration. Through our study, we not only document the existing practices that contributed to the project's success, but also propose improvements to further strengthen OSH design processes.


\subsection{Thingiverse}\label{thingiverse}

\begin{figure}[h]
  \centering
  \includegraphics[width=3.1in]{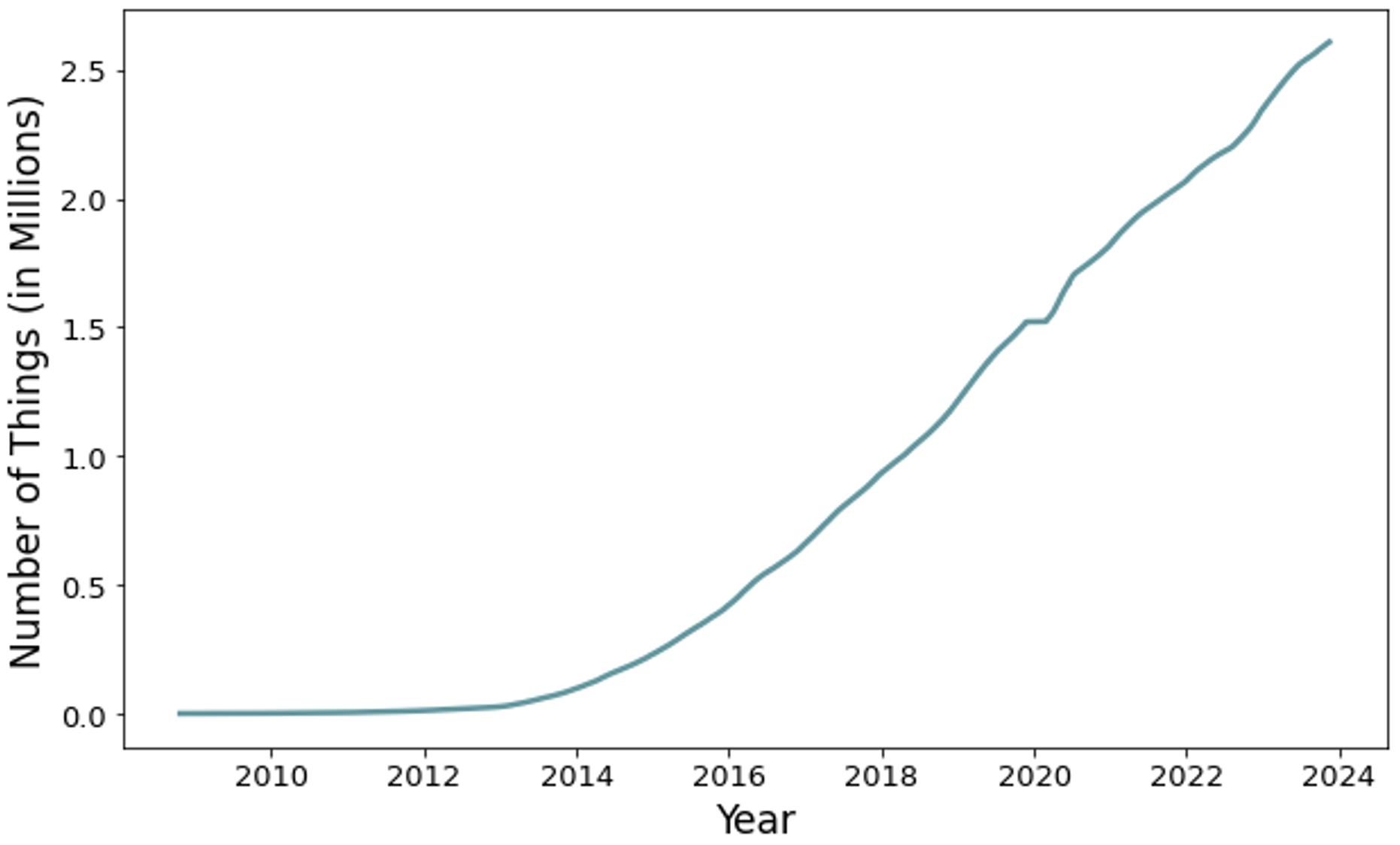}
  \caption{Cumulative number of Things published on Thingiverse from January 2009 to November 2023.}
  \Description{Number of Things published on Thingiverse from January 2009 to November 2023.}
  \label{thingiverse_growth}
\end{figure}

In this work, we focus on open-source hardware development on the Thingiverse platform. Thingiverse self-describes as a ``thriving design community for discovering, making, and sharing 3D-printable things,'' and is the largest online maker community for 3D CAD models~\cite{thingiversecom_thingiverse_2023,alcock_barriers_2016,kyriakou_knowledge_2017}. As of November 2023, Thingiverse contains over 2.5 million published designs~\cite{thingiversecom_thingiverse_2023}, with steady growth since its inception in January 2009 (Figure~\ref{thingiverse_growth}). Due to this popularity, Thingiverse has long been the platform of interest for researchers seeking to better understand OSH practices. In this section, we first briefly introduce several concepts of the platform, followed by a review of related Thingiverse literature.


\subsubsection{Background}
Thingiverse's primary purpose is to provide an open online repository of 3D design models, known as \textit{``Things''}~\cite{fordyce_things_2016}. Each Thing includes a variety of metadata: name, author's username, publication date, description (analogous to a README file), and uploaded files (e.g., CAD files, images, documentation) -- see Figure~\ref{drawbot_metadata} for a snapshot of a Thing page. The author must provide all of this metadata to publish the design. Users can interact with Things through likes, downloads, comments, makes, and remixes -- a \textit{``make''} refers to a user-uploaded photo or description showcasing their successful creation of a Thing, and a \textit{``remix''} is a user-generated Thing derived from one or more existing Things. 

Remixing resembles ``forking'' in software repositories, which allows makers to copy and modify a design to fit their needs, while formally establishing traceability between related designs, since the platform tracks remixing relationships. The majority of designs on Thingiverse are shared under open licenses (e.g., Creative Commons) that explicitly allow remixing adaptations~\cite{friesike_what_2021}.

\subsubsection{Related Work} 
This section provides an overview of relevant studies on design practices within the Thingiverse platform, covering how makers engage with Things, the creation of derivative Things (remixes), and the overall experiences of Thingiverse makers.  
Prior work has focused on scraping metadata of Things (e.g., likes, views) to understand the types of designs on Thingiverse~\cite{ozkil_collective_2017}, characteristics of popular designs~\cite{papadimitriou_remix_2015,papadimitriou_towards_2014}, and user engagement dynamics~\cite{novak_500_2020,novak_popular_2019}. 
Novak et al. conducted a longitudinal study on Thingiverse user interactions on Thingiverse, revealing that 
commenting is associated with experienced makers~\cite{novak_500_2020}, since commenting requires a deeper understanding of the design and motivation to actively engage with it~\cite{tan_conflict_2022}. 
Remixes were high-effort actions, requiring 3D CAD skills, creativity, and a considerable time investment~\cite{novak_500_2020}.   


\paragraph{Remixing}
Remixing plays a critical role in Thingiverse; in fact, more than half of Things are derived from prior Things~\cite{flath_copy_2017,ozkil_collective_2017,gillier_does_2020}. This significant practice has been of particular interest to HCI and CSCW researchers~\cite{stemasov_road_2021}.
Oehlberg et al. conducted an in-depth analysis of remixing behaviour in Thingiverse, to explore how remixing activity was impacted by \textit{Customizer}\footnote{http://thingiverse.com/apps/customizer} -- a Thingiverse-integrated built-in web app that allows simple modifications of 3D designs using pre-defined parameters~\cite{oehlberg_patterns_2015}.
Things created with Customizer did not elicit subsequent user activity, and the authors of these Things did not participate in experimental, creative, or collaborative projects~\cite{oehlberg_patterns_2015}. This finding is consistent with Flath et al.'s investigation of ``shallow'' (low degree of change) and ``deep'' (high degree of change) remixes~\cite{flath_copy_2017,wirth_patterns_2015}. 
Although these studies are enlightening, there are limitations with their coarse method of classifying ``deep'' or ``shallow'' remixes, and manual coding of remix relationships was deemed necessary for future work~\cite{flath_copy_2017}.

In addition to analyzing patterns of remixes, researchers have also explored the relationships between remixing and innovation~\cite{baik_role_2022,tan_conflict_2022,tan_empirical_2020}. When investigating the types of added value in remixed designs, Voigt found that remixes do not necessarily always contribute new design elements (referred to as ``feature-driven innovations''), but they can also streamline the design reuse process through linking related designs~\cite{acher_customization_2014}, or providing different file types of the same design (called ``production process-driven innovations'')~\cite{voigt_not_2018}.

While researchers have explored remixing patterns and innovation, they have predominantly focused on the incremental changes between remixed designs. What remains unexplored is how remixing occurs in a collaborative project, where contributions made in remixes can contribute to the overall evolution of a design project. Our study aims to address this gap and shed light on the collaborative contributions of remixing.




\paragraph{Makers}
While many researchers have studied the designs published on Thingiverse, others have focused on the designers themselves, understanding practices, such as design discovery~\cite{liang_customizar_2022}, license choices~\cite{moilanen_cultures_2014}, motivations for participation~\cite{shaw_why_2018,hausberg_why_2020,friesike_creativity_2019,stanko_disentangling_2022}, and how motivation impacts engagement~\cite{stanko_disentangling_2022,ozkil_collective_2017}. 
Makers can be active contributors to the design ecosystem, or more passive ``lurkers'' who observe and read other members' content without contributing their own~\cite{shaw_why_2018}. Seeking to comprehend the diverse user roles on Thingiverse, Li et al. studied user activity in a Thingiverse \textit{group}.\footnote{\textit{Groups} are self-governed sub-communities within Thingiverse where makers gather based on shared interests. Group members can discuss in forum posts, publish designs, and connect with other members of the group.} They found that the most active members contribute most to the group (in posts and likes), but
it is unclear whether roles would manifest differently in collaborative projects where members work towards shared design goals~\cite{li_dynamic_2023}.

Regardless of motivation, type, or activity, all makers will face challenges in the design and development process. Two pain points identified in existing work include tracking design versions and navigating the dependencies between related designs, which can pose obstacles to productive remixing practices~\cite{oehlberg_patterns_2015}. 
Alcock et al. identified pain points with understanding product functionality and customizability, proposing corresponding design recommendations for platform builders to develop beginner-friendly CAD tools~\cite{alcock_barriers_2016}.
While these studies provide insights into the pain points experienced by individual makers, we still lack an understanding of the challenges that impede productive collaborative design efforts.

Previous research has examined various aspects of open-source hardware on Thingiverse, such as published designs, remixing practices, and the experiences of makers. These studies provide large-scale overviews of metadata, thus offering insights into Thingiverse as a collective design platform. However, there is a lack of investigations of collaborative design on Thingiverse, which would require coordinated action and may face different challenges compared to collective design (see Section~\ref{OSHW}). As such, an exploratory study is needed to investigate collaborative design practices in a successful and well-maintained OSH project.


\section{Methods}

Our goal is to understand how makers collaborate in an OSH project within a collective design community. To do this, we conducted a detailed case study of designs hosted on Thingiverse. We chose a case study approach because it is the most suitable method for answering the ``how?'' question in a contemporary phenomenon~\cite{yin_case_2018}, such as OSH collaboration. Furthermore, case studies are the preferred method for exploring a phenomenon that is closely tied to its real-world context~\cite{yin_case_2018} -- in our case, a collaborative project situated within Thingiverse. Separating the design process from the context of Thingiverse is neither feasible nor desirable, as our aim is to comprehend how collaboration unfolds on a traditionally collective design platform. 

Our project-level analysis allows us to attain a comprehensive understanding of the design and development process from the project's inception. By doing so, we can glean insights into exactly how collaboration occurred -- answering questions such as who was involved, the duration of their involvement, the challenges encountered, and the contributions that resulted from the collaborative efforts. Focusing on one self-contained project allows us to grasp the nuances of collaborative processes, thus addressing our research goals. Thus, a case study of a single project is the ideal way to explore OSH collaboration practices. 

Before delving further into our methodology, it is important to explain our definition of a ``project''. OSS projects on GitHub are rather straightforward -- generally, a repository represents one project~\cite{Borges2016}. However, on Thingiverse, projects are not well-defined. Our case centres around one particular Thing posted on Thingiverse, but we expand the scope of the project to encompass the individual Thing as well as all related Things in ancestors and remixes. This decision was prompted by the platform's constraints, wherein only the original author of a Thing can modify it. Unlike in OSS, on Thingiverse, collaborators cannot contribute to the project through pull requests. However, Thingiverse's remix feature allows us to capture forking behaviour, to an extent. Thus, while we are interested in one particular Thing, we considered all of the ancestor Things, as well as all remixed Things, as part of the complete project. To the best of our knowledge, our study is the first to consider this network of designs as a project on Thingiverse.

\begin{figure}[h]
  \centering
  \includegraphics[width=5.1in]{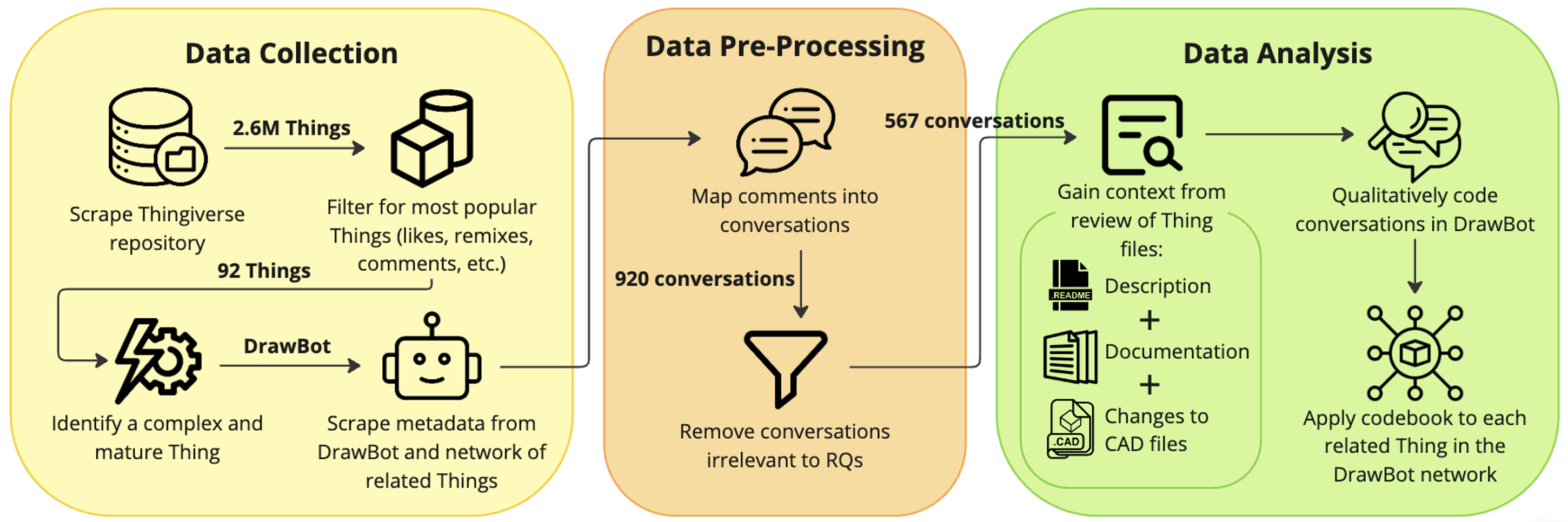}
  \caption{Overview of research methods, from data collection to analysis.}
  \Description{Overview of research methods, from data collection to analysis}
  \label{methods}
\end{figure}

In this section, we first provide a rationale for our project selection, followed by an overview of our data collection, processing, and analysis methods. Figure~\ref{methods} summarizes our research method.

\subsection{Project Selection}\label{project_selection}


In this study, the project we focus on is the \textit{``Drawing Robot - Arduino Uno + CNC Shield + GRBL''}.\footnote{https://thingiverse.com/thing:2349232} 
The Drawing Robot -- or DrawBot -- is a pen plotter device designed to create detailed 2-dimensional illustrations from pre-programmed sequences of movements provided by the user. Our selection of the DrawBot for this case study followed the steps outlined in Figure~\ref{methods} and is summarized below. 

First, we aimed to select an active and popular project, with high engagement among Thingiverse users. To do this, we filtered Things to obtain the 99th percentile in engagement metrics (i.e., likes, views, and downloads) and metrics that signify high-quality designs (i.e., number of comments, threads, makes, files, remixes, and ancestors); in total, 92 Things were in the 99th percentile across all nine metrics. 
\textcolor{black}{We explain how to interpret these metrics below:}
\begin{itemize}
    \item \textit{Likes, views, and downloads:} Higher numbers signify more activity without known tradeoffs in innovation or complexity.
    \item \textit{Comments and makes:} Higher numbers signify more activity, but comments suggest experienced makers~\cite{novak_500_2020,alcock_barriers_2016} -- so, we prioritize a high comment and thread count. 
    \item \textit{Remixes:} A higher count indicates a more popular Thing, but we avoid heavily remixed Things (100+ remixes) that might signify mass customization with low complexity~\cite{Blikstein2013,voigt_not_2018}. 
    \item \textit{Ancestors:} Due to the inverted U-shaped relationship between the number of ancestors and innovativeness~\cite{baik_role_2022,tan_conflict_2022}, we avoid Things with no ancestors, and Things with many ancestors.
\end{itemize}
Overall, these descriptive statistics help us to maximize a Thing's activity and innovativeness while avoiding potential pitfalls associated with projects scoring highest on these metrics.
After obtaining 92 things that met our filtering criteria, we manually inspected them for suitability based on the following considerations: product complexity and project maturity.




The chosen product needed to be complex to offer us a rich and comprehensive investigation of design collaboration. Consequently, complexity comprised three criteria: (1) the product must serve a practical purpose and function to avoid trivial items (e.g., figurines); (2) the product must be an assembly -- products consisting of multiple parts that interface and interact~\cite{Noort2002};
and (3) the product should be electro-mechanical (i.e., incorporate both hardware and software components). In recent years, electronic toolkits (e.g., Arduninos) have become a staple in many open hardware projects~\cite{naik_modularity_2021,mellis_collaboration_2012,stemasov_design_2023,desai_assembly-aware_2018}, allowing makers to create software-controlled devices (e.g., robots)~\cite{Cvijikj2011}. These electro-mechanical components require makers to design and configure both hardware and software, and are associated with more complex products~\cite{jacobs_product_2007}.

Lastly, the project must have been published for at least several years (we arbitrarily selected 4), since open-source hardware projects often require \textit{``several years [to] indicate success''}~\cite{antoniou_defining_2022}, with long-term sustained active involvement signalling a well-maintained project. Simultaneously, the project should still be active (i.e., growing in comments and remixes), reflecting ongoing refinement. However, a slowing growth in activity signals that the project has entered a mature stage, with the primary development work likely completed. By selecting a project that has been active over several years, we can observe a more in-depth understanding of the collaborative design process. 




In summary, our selection process ensured the Drawbot's high user engagement, sufficient complexity, and long-term sustained activity. 

\subsection{Data Collection and Pre-Processing}\label{data_collection}
We used Thingiverse's public API to extract metadata from all published Things, totalling 2,608,691 as of November 2023. Scraping all Things was necessary to filter projects for our selection process.

\subsubsection{Collecting metadata for each Thing. }
\begin{figure}
  \centering
  \includegraphics[width=4.9in]{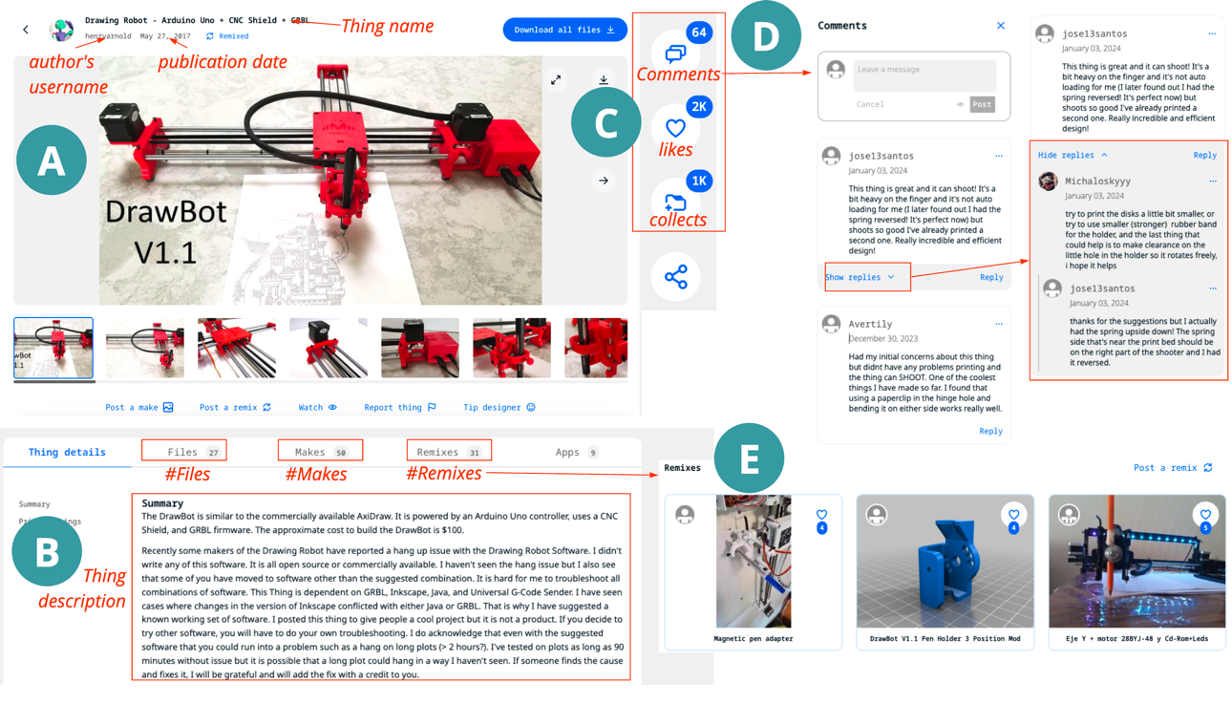}
  \caption{Drawing Robot ``Thing'' studied in this paper. The metadata includes: (A) Thing name, author's username, and publication date; (B) Thing description, number of files, makes, and remixes; (C) user activity (e.g., likes, comments, collects); (D) the comment panel, with commenter usernames, posted dates, and threaded replies; (E) the remix panel, including remix activity with page links and author usernames.}
  \Description{Screenshot of the Drawing Robot ``Thing'' studied in this paper.}
  \label{drawbot_metadata}
\end{figure}

\begin{figure}[t]
  \centering
  \includegraphics[width=3.8in]{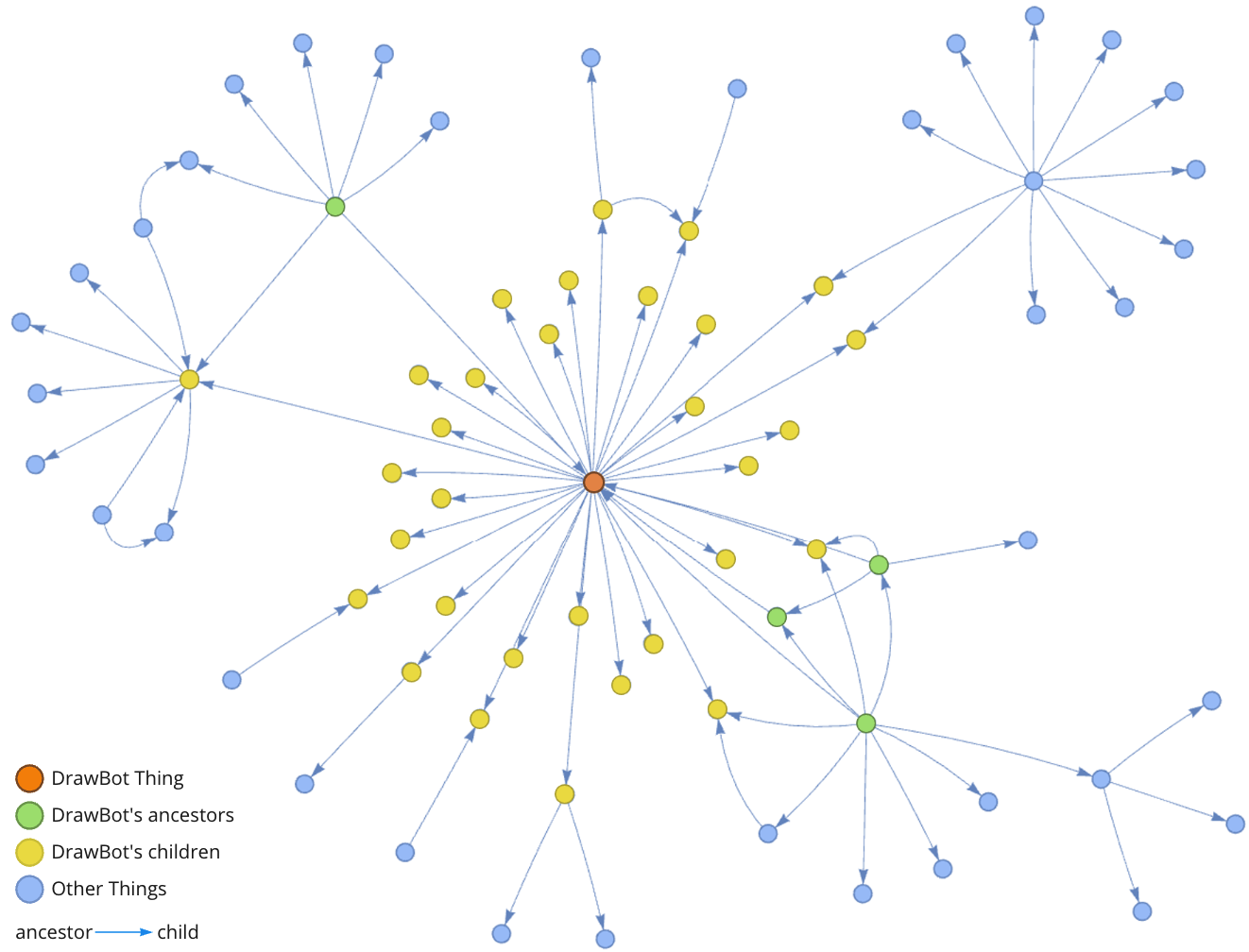}
  \caption{Directed network of Things within the DrawBot project. The direction of the arrows indicates a remix relationship, pointing from ancestor Thing to child Thing. The orange node represents the DrawBot Thing; yellow nodes represent DrawBot remixes; green nodes represent DrawBot ancestors; and blue nodes represent other Things within the network indirectly connected to the DrawBot Thing (i.e., 2 or more nodes away).}
  \Description{Directed network of Things within the DrawBot project.}
  \label{drawbot_network}
\end{figure}

Once the DrawBot project was selected, we used the Thingiverse API to scrape the relevant metadata presented in the DrawBot Thing page as shown in Figure~\ref{drawbot_metadata}. We also collected all metadata for each Thing in the project (DrawBot's ancestors and children). Figure~\ref{drawbot_network} provides a network of the 74 Things in the DrawBot project, where nodes are Things and edges are remixing relationships (pointing from parent Thing to child Thing);
four Things are direct ancestors of the DrawBot, 31 Things are direct remixes from the DrawBot, and the remaining 38 Things are either ancestors or children of the directly related Things. 
Although we collected all of this metadata, our analysis focused on the comments, descriptions, and posted files, which typically included 3D CAD models and documentation. \textcolor{black}{We have made our full dataset publicly available for use by the research community~\cite{cheng_data_2024}.}

Of the 74 related Things in the DrawBot project, 31 Things had comments, totalling 2,483 comments. The average length of a comment was 48 words (SD = 58). Across 74 Things, there were 50 unique authors and 684 unique commenters. On average, each author made 19 comments (SD = 92), and non-author commenters made 2.3 comments (SD = 3.0). The average length of a description was 168 words (SD = 392); the DrawBot description itself was the longest, at 3,179 words. The average file count was 5 (SD = 5).

\subsubsection{Collecting and preprocessing comments.}
In order to understand discussions among DrawBot contributors, we collected comments from each Thing (component D in Figure~\ref{drawbot_metadata}). Note that the comment panel lists comments linearly in chronological order. Users could reply to a comment to have a cohesive conversation. Thus, we iteratively collect all the comments and nested replies.

\paragraph{Constructing discussion threads.} 
After collecting all comments, we read through each comment chronologically to map conversations together. Although Thingiverse has a threaded reply feature, it was not used consistently, necessitating an initial read-through. We define a conversation as an exchange between multiple people who are discussing the same topic; for example, in Figure~\ref{drawbot_metadata}, two makers are discussing a troubleshooting topic and replying directly to each other.
In total, there were 920 conversations. Throughout this process, we also tagged conversations for relevance to our research questions; conversations solely expressing praise or appreciation, such as \textit{``I love your machine and your artwork''} or \textit{``Thank you so much for your time,''} were deemed irrelevant and discarded; of the 920 conversations, 353 were discarded. In the remaining 567 relevant discussions, there were 402 unique commenters.

\subsection{Qualitative Data Analysis}

Following data processing, we began the qualitative content analysis~\cite{krippendorff_content_2018} in order to determine how designers collaborated in the project. \textcolor{black}{One coder conducted the analysis.}

After an initial examination of the DrawBot Thing description and documentation to understand the design and function of the device, the first author independently coded the conversations of the DrawBot Thing (567 conversations, comprising 1,540 comments) using a bottom-up coding approach~\cite{Saldana2009,Creswell2014}, because there is no widely accepted categorization of collaborative contributions in OSH, and the exploratory nature of our research questions. For each conversation, they coded: (1) the component of the DrawBot to which the conversation pertained, and (2) the collaborative contribution made by the commenter in the conversation.

To guide the coding of the component, they referred to a comprehensive parts list provided by the DrawBot author. This document outlined all the necessary components for constructing and operating the DrawBot, categorizing them into hardware, electronics, or software. Our codebook for relevant components was structured based on this detailed documentation (see Table~\ref{codebook} in Appendix~\ref{appendix_codebook}). 

In addition to the project component, \textcolor{black}{the first author} also coded the task accomplished in the conversation. \textcolor{black}{They} focused on discussions resulting in productive outputs or the generation of new knowledge.
Many codes emerged from commenters suggesting ways to improve various DrawBot components; others provided troubleshooting help for makers struggling with printing or building their DrawBot; and some codes were developed because makers identified problems with the design or fit of parts. 

After the first round of open coding, a second round was conducted to ensure consistency between comments analyzed at the beginning and end of the process. A third round involved axial coding to group similar codes, identifying higher-level themes. This process required iteration between the comment data and related literature to refine the theme categorization, which is characteristic of qualitative research analysis~\cite{Corbin2008}. The themes categorized tasks into \textit{design-related} and \textit{non-design-related} contributions.

Once the codebook had been established, a fourth and final round of coding applied the established codebook to other related Things in the DrawBot project (200 conversations, comprising 380 comments). Similar to our process of coding comments of the DrawBot Thing, we also read the Things' descriptions, analyzed their attached files, and used CAD software to compare versions of the same CAD models, to understand the context of these comments. 





\section{Results}
In this section, we present the findings of the DrawBot case study. First, we provide an overview of the collaborative design activity in the project. Next, we elaborate on how makers collaborated on both design-related and non-design-related tasks. Lastly, we present the collaboration challenges faced and how the community overcame them. As per ethical guidelines for internet research~\cite{franzke_internet_2020}, we have paraphrased all quotes used in this paper to protect the anonymity and searchability of Thingiverse users.


\subsection{Collaboration Practices (RQ1)}\label{collab_practices}
In this section, we provide a detailed account of collaboration practices in the DrawBot project, shedding light on the types of tasks requiring collaboration (RQ1.1) and how makers collaborated to complete them (RQ1.2).
Through analyzing comments and remixes, we found that makers collaborated to fulfil: \textit{design-related} and \textit{non-design-related} contributions (summarized in Table~\ref{comment_types}). 
In the subsequent sections, we describe collaboration scenarios that occurred in the project.

\begin{table}[ht]
\footnotesize
  \caption{Types of collaborative contributions found in the comments. Contributions to open-source hardware collaboration can be \textit{design-related} or \textit{non-design-related}. We also report the percentage of conversations that discussed the contributions. Note that the percentages do not add to 100\% since one conversation could discuss many topics.}
  \begin{center}
  \label{comment_types}
  \begin{tabular}{p{.6in} p{.7in}>{\raggedright\arraybackslash} p{1.6in} p{1.6in} c}
      \toprule
    Category&Sub-category&Description&Example(s) from Dataset&\%\\
    \midrule
    Design-Related&New Feature&Designing and developing new functionality for the design.&\textit{``I am adapting DrawBot to draw with 12 colors.''}&10.8\\ 
    &Error Fix&Modifying components to correct issues or errors in the design.&\textit{``You need to modify the corrupted line in the MI GRBL Extension.''}&4.3\\ 
    &Performance Enhancement&Optimizing the design to improve efficiency (e.g., power, speed).&\textit{``You could lower the mass so that the drawing robot can move faster.''}&2.8\\ 
    &Component Replacement& Substituting elements with alternative components.&\textit{``Could a Raspberry Pi be used in place of an Arduino?''}&6.2\\
    \midrule
    Non-Design-Related&Issue Tracking \& Management&Reporting, updating, and requesting support for issues.&\textit{``New error as of Dec. 2019: Users are getting Python errors.''}&10.6\\
    &Documentation&Writing and improving documentation.&\textit{``I will change the instructions to point to the better MI extension.''}&5.3\\
    &Version Tracking&Reporting and managing changes to the design.&\textit{``I posted an updated top clamshell. Make sure you have the latest.''}&1.3\\
    &Community Support & Participating in discussions, providing troubleshooting support, and helping others.&\textit{``How can I help you? How are the parts breaking, or do you mean the whole machine does not work?''}&66.0\\
  \midrule
\end{tabular}
\end{center}
\end{table}

\subsubsection{Overview of the DrawBot Project}


Over the course of the DrawBot project, collaborations between designers resulted in numerous contributions that evolved the DrawBot design. Through qualitative coding, we were able to identify which of the 402 commenters actually made design changes to the DrawBot. Furthermore, we are able to obtain a detailed understanding of the topic, duration, and outcome of the collaboration. 

\begin{figure}[h]
  \centering
  \includegraphics[width=4.7in]{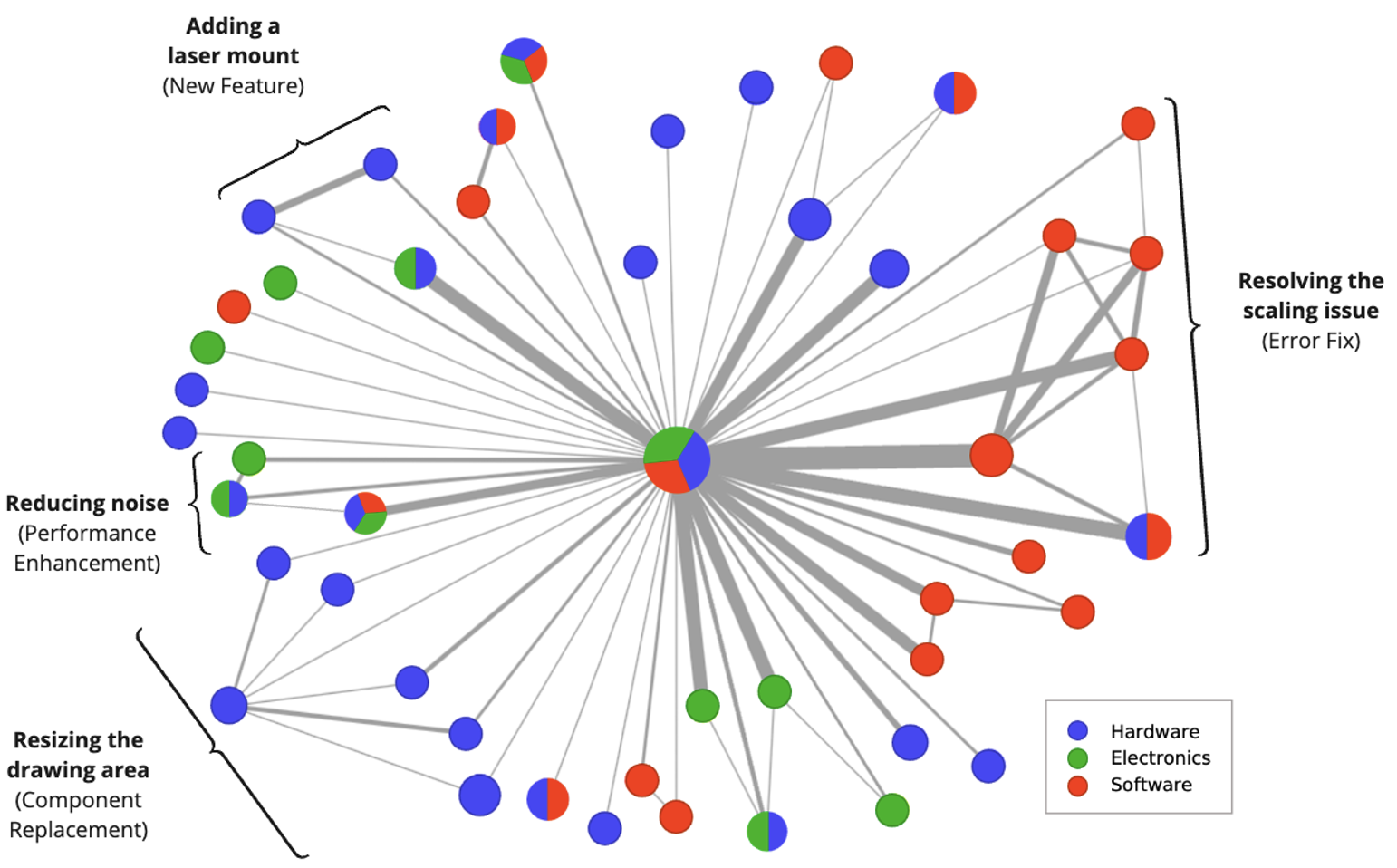}
  \caption{Contributors to the DrawBot project. Each node represents a maker who made a design-related contribution, with the size of the node representing the number of contributions made (note: for the sake of figure legibility, we scaled down the central node since this was the DrawBot author, who made numerous contributions). The edges represent conversations (i.e., threads) shared between makers, with the edge weights showing the frequency of collaboration (i.e., the number of conversations). Blue, green, and red nodes represent exclusively hardware, electronics, and software contributions, respectively. Multi-coloured nodes represent interdisciplinary contributors.}
  \Description{Collaboration network of DrawBot contributors.}
  \label{collaboration_network}
\end{figure}

The network graph in Figure~\ref{collaboration_network} summarizes the collaboration activity among all 48 makers who made design-related contributions to the Drawbot; we call them ``contributors''. This network shows which contributors collaborated, the extent of the collaboration, the specific components of the DrawBot they discussed, and the contribution made (whether hardware, electronics, software or a combination). In particular, there are four highlighted clusters in the network, which each represent an instance of collaboration that we will describe in Section~\ref{design_related}. 


\subsubsection{Design-Related Contributions}\label{design_related}
\begin{figure}[ht]
  \centering
  \includegraphics[width=4.2in]{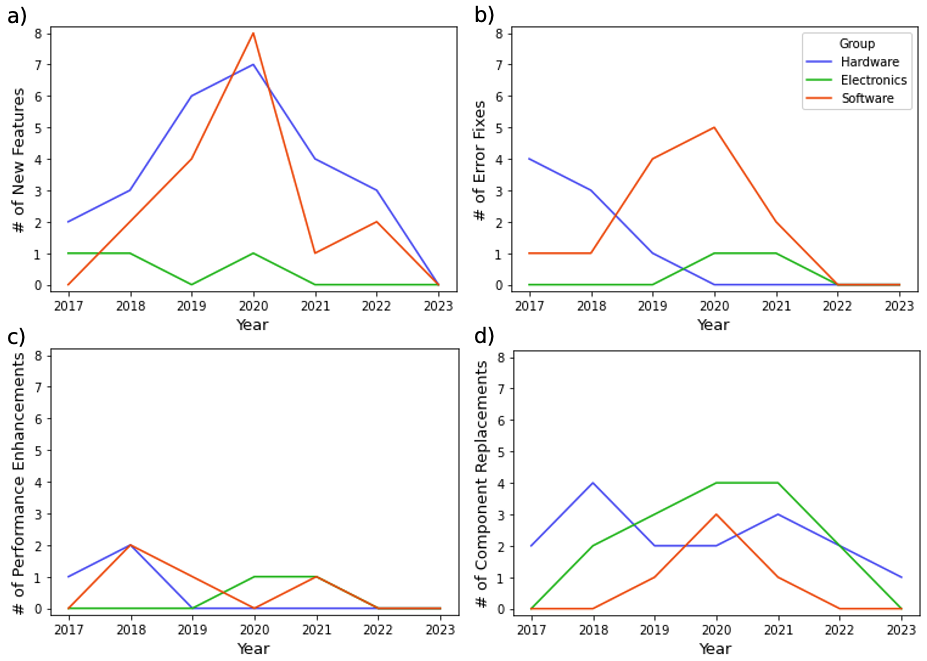}
  \caption{Number of design-related contributions made to the DrawBot's hardware, electronics, and software components over time, based on contribution type: (a) new features; (b) error fixes; (c) performance enhancements; and (d) component replacements.}
  \Description{Characteristics of the Thingiverse data set.}
  \label{changes_over_time}
\end{figure}

Design-related contributions pertain to the actual DrawBot device. The examples presented in this section describe how makers collaborated to modify or improve one or more of the components of the DrawBot, which may include changes to the CAD models, code files, configuration of the electronics, or assembly. To give a better idea of when and how often these contributions were made to the project, Figure~\ref{changes_over_time} shows the number of design-related contributions per year of the project's active development.

\paragraph{\textbf{New Feature}}
The most common design-related contributions were \textit{New Features}, encompassing additional functions, parts, or enhanced features of existing parts within the DrawBot. As shown in Figure~\ref{changes_over_time}a, the number of hardware and software contributions peaked in 2020, and there were few new electronics features overall. Of the relevant conversations, 10.8\% discussed adding a new feature to the design. 

Many makers were interested in adding parts to the pen holder component to hold items beyond pens, such as pencils, vinyl cutters, and paintbrushes. In particular, three makers collaborated to add a laser mount, enabling wood engraving. The author initially proposed this idea in June 2017, stating, \textit{``I intend to experiment with a laser mount.''} At first, uptake on this task was slow, until newcomers to the project were also interested in the laser functionality. The author responded to each new collaborator with the information they had, such as, \textit{``Check out this extension, which provides guidance on connecting the laser controls to the Arduino board. I haven't tested it but it looks promising.''} As the collaborators made progress in designing, prototyping, and testing the new design, they reported updates in the comments, like, \textit{``I needed to replace the rods to support the weight of the laser module''.}

Eventually, in December 2019, one of the collaborators responded with a final solution, stating, \textit{``I used the Neje 3500 laser, which is compatible with the existing firmware. My first attempt at a custom mount is working well so far.''} To which the author responded, \textit{``Great info! I'm planning to do this too. Please share your STL files as a remix.''} Unfortunately, the profile of the maker who contributed to the new laser mount has since been deleted, so we were not able to confirm if the files were posted. Nonetheless, they contributed new knowledge as to how to incorporate a laser into the DrawBot.

\paragraph{\textbf{Error Fix}}\label{error_fix}
Error fix contributions were necessary when there was a problem with a component that either rendered the DrawBot unusable or negatively impacted the DrawBot's manufacture or functionality. Error fixes were mentioned in 4.3\% of conversations. 

The first design-related contribution to the DrawBot, aside from the initial post by the author, was a hardware error fix. This fix involved redesigning the clamshell parts after another maker faced difficulties with printing and assembling, mentioning, \textit{``I could force fit the bearings into the clamshell, but the two clamshell halves are still misaligned, causing severe bending of the rods''.} After the author uploaded an updated clamshell model, a new error was introduced, this time in the 3D printing of the part, with another maker reporting, \textit{``still struggling with the clamshell print, it keeps lifting off the base. Could you check if the clamshell is properly flush with the print base?''}; to which the author replied, \textit{I apologize, you were right! The top clamshell had a slight 0.2-degree angle, which I overlooked. Please see Revision 3 of the file.''}

As shown in Figure~\ref{changes_over_time}b, most hardware error fixes occurred early in the DrawBot Thing's publication, gradually decreasing until 2020, when no further error fixes occurred. This trend could be the case because hardware errors prevent the maker from being able to successfully print and assemble the device -- which is a required prerequisite to wiring electronics and programming software -- so, problems with the 3D models were quickly identified and resolved. Furthermore, since a maker would typically only make a DrawBot once, and physical parts remain unchanged (unless they require replacement or upgrades), a more mature project sees fewer hardware errors to fix, as DrawBots are already up and running smoothly.

Although hardware error fixes decreased over time, software errors exhibited the opposite trend, steadily increasing until 2020. In January 2020, the DrawBot community faced a significant software issue with a compatibility bug between Inkscape and the MI Extension, resulting in scaling problems with drawings. One maker identified the error, \textit{``Hello everyone, has anyone encountered the same problem I'm facing with Inkscape's scaling of G-code output?''}, followed by many other makers reporting the same problem. 

Initially, the author proposed a fix suggesting adjustments to the MI extension's Python code for G-code creation, stating \textit{``Try this workaround: adjust the drawing scale to 3.5433''}, which worked for some types of illustrations (e.g., text writing) but not all. 
After two months of employing this workaround, a maker decided that it was time for a more permanent and holistic solution, stating, \textit{``I wanted to address the scaling issue that 
to this point has remained unresolved. I've been looking into this, and here's what I found''}, accompanied by an extensive list of links to online resources.

This launched a trial-and-error process involving six makers who posted their ideas, shared online resources, and tested each other's solutions on their DrawBots. After several discussions until August 2020, a fix was finally found for the original MI Extension software, but the six makers simultaneously became interested in replacing the extension with a newer, more reliable version. This group of six makers again dedicated themselves to resource sharing and testing, leading to a collaborator presenting a fix in a remix in November 2020. As described by the collaborator, \textit{``I created an updated Inkscape Extension that fixes problems with the original, and is also compatible with the latest version of Inkscape.''}

\paragraph{\textbf{Performance Enhancement}}
Over the course of the DrawBot project, performance enhancement contributions were made the least, mentioned in 2.8\% of conversations. One priority was to increase DrawBot's drawing speed. 
One collaborator experimented with modifying hardware components: \textit{`` I've replaced the stainless steel rod with 8 mm aluminium tubes to reduce mass, enabling the drawing robot to move faster.''} Once they posted a successful remix of this design change, this inspired another maker to follow suit, responding to the collaborator who posted the remix, \textit{``I found the rods in a nearby shop! They're cheaper and easier to cut.''}

In the context of this 3D printing community, makers may find it relatively straightforward to enhance the performance of hardware components (e.g., designing lighter parts). However, when it came to electronic components, which are typically standardized and purchased off-the-shelf, makers rarely suggested performance enhancements. In a scarce number of instances where collaborators did discuss improving the performance of electronic components (e.g., CNC Shield), it was most often to suggest a replacement, rather than to offer new approaches to wire or solder the part. Over the month of February 2021, four makers collaborated in selecting replacements by providing links to various stepper drivers and discussing the advantages and disadvantages of each. After identifying the optimal one and successfully testing it, one of the collaborators suggested, \textit{``for a significant noise reduction, switch from A4988 to TMC2209 stepper drivers. Be aware that the TMC2209 drivers come with a larger heatsink that is not compatible with the provided housing in the Thing Files.''} Not only did they offer a replacement that would improve the noise reduction, but they also detailed necessary hardware adjustments to accommodate the new drivers, highlighting a holistic approach to performance enhancement.

As shown in Figure~\ref{changes_over_time}c, performance enhancements were scarce, which aligns with existing literature that suggests OSH communities often emphasize creating alternative designs rather than continuously refining a single central design~\cite{mellis_collaboration_2012}. This trend may be attributed to a lack of engineering expertise in design optimization or a lack of incentive to pursue the ideal design, given the maker-oriented nature of Thingiverse. 

\paragraph{\textbf{Component Replacement}}
Although design-related contributions are typically associated with the DrawBot's enhanced performance, or the introduction of a new feature, we found that some contributions do not necessarily improve the design. Rather, 6.2\% of conversations discussed replacing a component in the design. Design changes as a result of \textit{Component Replacement} happened most often because the maker already had a particular component and wanted to make use of it. For example, one commenter asked, \textit{``Can the CNC shield v4 be operated with an Arduino NANO? I have a spare one.''} 
Another common reason was the unavailability of specific materials in the maker's geographic area. For example, one maker expressed, \textit{``I can't find Nema 17 0.4 amp motors in my country. Would it be fine to use a different Nema 17 motor and modify the driver current?''} 


Beyond the lack of component availability, some makers also cited material tradeoffs as a reason for \textit{Component Replacement}. Even though the open-source nature of the project implies technical freedom in design, the reality of creating a physical product requires the end user to procure the hardware materials. Thus, there exists an inherent tradeoff in the selection of materials, often influenced by considerations such as component costs. For instance, in June 2018, one collaborator posted a remix, saying, \textit{``Here's a budget-friendly version. I cut costs by swapping the rods with hollow, shorter ones and the metal bearings with 3D printed ones.''} 

After this collaborator successfully downsized their DrawBot's drawing area by replacing the rods, other makers followed suit, referring to the original maker who posted the change. In total, five separate makers consulted with this maker to achieve the component replacement task. This cluster of makers is shown in Figure~\ref{collaboration_network}. Since the discussion on rod replacement involved fewer iterations and less problem-solving, the five nodes are not interconnected but rather directly connected to the first maker who successfully achieved the dimensions.


In each of these examples, the primary motivation behind the design change was working around practical obstacles. Obstacles related to component availability, cost, or ease of assembly are key considerations in the design of physical products. Figure~\ref{changes_over_time}d shows that the occurrence of component replacements remained relatively constant over the project's duration. This could be attributed to the nature of component replacements being driven by individual makers' needs (e.g., the materials available to them), and this lack of access is not affected by the evolution of the design.

\subsubsection{Non-Design-Related Contributions}
Although non-design-related contributions do not involve changing the actual DrawBot, they are essential for driving the development work. These non-design-related contributions are essentially ``articulation work,'' or \textit{``work to make the cooperative work work''}~\cite{schmidt_taking_1996}. In this section, we highlight how non-design-related contributions facilitated streamlined collaboration.

\paragraph{\textbf{Issue Tracking \& Management}}
Throughout the project, makers used comments to raise and discuss issues in the comments, encompassing both problems and tasks for the community to address. 10.6\% of the conversations were about tracking and managing issues.

The issue tracking process typically followed three stages: (1) reporting (e.g., \textit{``FYI, the GRBL parameters displayed in the images and in the PDF instructions are different''}); (2) acknowledging or undertaking the task (e.g., \textit{``Yes, I'm aware of some discrepancies in the documentation. I plan to fix these in the next revision''}); and finally, (3) resolving or updating (e.g., \textit{``I corrected links in the documentation and made minor edits.''}).

Issues were not always in response to problems; sometimes, an issue signified a task or specific area in which the community should focus their development efforts. For instance, when aiming to expand the software capabilities for different types of illustrations, the author issued an open call for future directions, stating, \textit{``If you're open to enhancing the design, consider upgrading the software to a newer version of GRBL for added features. Additionally, explore a controller that integrates Arduino and CNC circuitry, which can also support a laser. Understandably, if it's too much, the current design works well, but improvements could be exciting.''}

\paragraph{\textbf{Documentation}}
Contributions relating to documentation involved the creation of new documentation or the improvement of existing documentation, which included user manuals, images, and assembly instructions. Documentation contributions represented 5.3\% of conversations.

In the initial publishing of the DrawBot, the author provided a parts list, CAD files necessary to print the DrawBot, and links to software downloads with screenshots of correct parameters. However, there was no user guide with full manufacturing and assembly instructions, until one enthusiastic maker created and shared an 84-page guide on Google Slides: \textit{``Here's the link to an instruction manual I made.''} Appreciative for the effort, the author responded, \textit{``Wow, impressive manual! Thank you! Can I share it in the Thing files? Otherwise, it might be overlooked in the comments. Feel free to include it in a remix of this drawing robot.''} As changes were made to the design, both the author and the creator of the documentation updated this user guide on Google Slides, and exported it as a PDF to include in the Thing files.

\paragraph{\textbf{Version Tracking}}
In the absence of sophisticated version control tools, the DrawBot community used the comments section to facilitate the tracking of design versions, notifying others about the availability of new design files.  Only 1.3\% of conversations contributed to version tracking.

One maker, aiming to enhance the pen holder design, created a remix and shared the updated design in the comments, stating, \textit{``Here is a link to a significantly improved pen holder. I'm creating one for my DrawBot.''} When the new holder was successfully printed and integrated into the DrawBot assembly, the original author then ``merged'' the new version by adding the CAD files from the remix into the original DrawBot repository. 

Not all design changes made in remixes were merged; only those considered significant by the author were included on the DrawBot page. Factors such as the severity and prevalence of the problem addressed by the new design influenced the decision. For instance, an improved version of a widely-used component, like the pen holder, was more likely to be merged, while the addition of a wall mount was somewhat of a diversion from the main design. These designs could still be found in remixes, but were not added to the Thing files.

\paragraph{\textbf{Community Support}}
Community support involved answering questions and providing troubleshooting support. Notably, 66.0\% of the comments in the DrawBot community were related to community support. This contribution involved answering various questions regarding: (1) 3D printing (e.g., \textit{``can I print these using PLA instead of ABS?''}); (2) the cost and procurement of materials (e.g., \textit{``filament aside, how much are the materials?''}); and (3) assembly (e.g., \textit{``Can someone show the wiring of the servo to the CNC Shield?''}). 

Providing troubleshooting help was another popular contribution. Makers seeking troubleshooting help often shared images and videos for hardware or electronics issues and provided their code and error messages for software-related problems. The DrawBot author promptly responded to all troubleshooting requests, offering supportive messages, such as, \textit{``Simply shift the servo up by one spline, and you should be fine. Don't be too hard on yourself. This project has a lot of moving parts, quite literally!''} 

As the project progressed, and more makers successfully printed and operated the DrawBot, they actively engaged in troubleshooting tasks to assist the original author. In one example, a maker was having issues fitting the belt into the clamshell, which prompted a contributor (who was not the author) to offer methods to tighten the belt. In acknowledgement, the author replied, \textit{``First check out the valuable suggestions provided by [Maker X] in the comment below. If that doesn't work, I have a few other ideas.''}

\subsection{Collaboration Challenges \& Workarounds (RQ2)}
Our objective with this study was to explore how collaboration unfolds in a successful OSH project. Although the DrawBot project was a success, there were challenges that impeded the collaboration's full potential. In this section, we present three collaboration challenges articulated by makers in the comment threads, alongside a discussion of how the DrawBot community overcame these challenges.

\subsubsection{Poor Version Control Support}\label{version_control}
On Thingiverse, the Thing page is essentially a snapshot of the repository, showing only the most recent version, rather than an overview of the entire design process~\cite{ludwig_towards_2014}. Since there is no version control support, some makers unintentionally used outdated components. In the clamshell scenario described in Section~\ref{error_fix}, despite a new CAD file being uploaded, four different makers were unaware of the update and downloaded and printed the outdated version, leading to issues highlighted in the comments, like \textit{``Is there an issue with the top clamshell? I've had difficulties printing it and problems with adherence.''} These situations not only frustrate makers but also result in wasted time and print material.

To address the lack of version control support and work around this challenge, DrawBot contributors had two strategies: remixing, and tracking changes in the description.


\paragraph{Remixing}
Remixing allows makers to copy and upload new versions of Things. In order to post a remix, it must contain a CAD file, which was straightforward for the majority of the DrawBot remixes which contained upgraded or new 3D models. However, this can be inconvenient for revisions to a software component. For instance, one maker replaced the GRBL firmware with a newer, more versatile system, and was asked by the author, \textit{``Please share a remix of your solution. You just need to upload one STL file along with information directing to your GitHub link.''} Employing this workaround, the maker copied a random CAD file from the DrawBot into their remix, but this is not an optimal solution. Future development is needed to support remixing of electro-mechanical designs in OSH.

\paragraph{Updating the Thing Description}
On a Thing page, the description section is intended to summarize a Thing's function and instructions for 3D printing and assembly. In addition to providing this information, we found that DrawBot's author used the description to track design changes and document version history. After each new revision, the author added a short summarization of the change with the revision date, similar to \textit{commit messages} in collaborative software development~\cite{tian_what_2022}. One such example is, \textit{``Update Dec 1, 2020: I've uploaded upgraded parts for the pen slider, pen holder, cable support anchor point, and Arduino enclosure mount, all of which are backward compatible with the originals. These enhancements result in a neater and sturdier DrawBot.''} While this process is manual and tedious, it currently serves as the only means to track and display revision updates on the Thing page.

\subsubsection{Scattered Collaboration}\label{scattered_collaboration}
Collaboration on Thingiverse can be scattered, which makes it difficult for contributors to maintain an awareness of the people, design artifacts, and documentation of the project.

Remixing is considered a way to maintain links between designs, but some Thingiverse makers suggest that remixing can actually harm collaboration through community fragmentation. When one maker requested an updated design for the motor mounts, they decided against creating a remix, and instead asked the author, \textit{``Could you please fix these clamshell holes? It would be a shame to do a remix and split the community''.} 

Maintaining an awareness of the contributions made within remixes is also challenging, especially when there are many remixes that span multiple pages that the user has to click through. Thus, it is common that even when someone takes the proper steps to create a remix, it can be overlooked by other makers. In the example where makers wanted to resize the drawing area, one maker pointed another to an existing solution, stating, \textit{``Other users have successfully enlarged the design and shared their builds. Check out their remix''}, accompanied by a link to the remix. 

Documentation can also be scattered, since many makers link to external image- or video-sharing platforms, like Imgur, YouTube, or Google Drive. For example, makers created YouTube videos to demonstrate assembly instructions. As stated by one maker, \textit{``I successfully configured and optimized my DrawBot. Check out the results in this video. Feel free to leave comments here or on YouTube if you have any questions.''} These uploads can attract makers not originally from Thingiverse to become involved in the project. However, the lack of integration between these platforms and Thingiverse further perpetuates the fragmentation of the community.




\subsubsection{Lost Contributions}\label{lost_contributions}
Although makers were encouraged to make remixes to share their design contributions, not all were captured in this method; many were provided in the comments instead. Currently, individuals other than the original author cannot directly modify the Thing, its metadata, or related files. Consequently, valuable fixes provided in the comments may be hidden amidst numerous conversations. To avoid this challenge, the author suggested, \textit{``Remix the DrawBot with your code as a separate file to prevent it from getting buried in the comments. With your permission, I may eventually incorporate your fix into the documentation, but for now, a remix would be greatly appreciated. I'll link to your remix from my Thing description.''} This example highlights two approaches to prevent lost contributions: creating a remix, and updating the Thing description (Section~\ref{scattered_collaboration}).

However, due to the subjective nature of what makers consider as derivatives and the voluntary nature of attributing credit in remixes, makers were not always disciplined in linking remix relationships. In the comments, one maker might direct another to a related Thing page, but that Thing is in no way linked to the original DrawBot page. As a result, makers would be unaware of these contributions unless they read the comments in detail.

While these challenges result from a lack of awareness of derivatives (i.e., downstream Things), contributions in sources (i.e., upstream Things) can similarly be lost. When makers were not properly given credit where credit was due, tensions arose in the community, as illustrated by a comment in one of the DrawBot network Things: \textit{``I created the original design. They simply `borrowed' it, modified the base, and failed to mention that they aren't the author of this plotter''}.


\paragraph{Crediting Contributors}
To avoid tensions, crediting fellow makers for their contributions is a simple, yet effective approach. The DrawBot author diligently followed this practice, and acknowledged collaborators in both the comments and Thing description, with quotes like \textit{``[Maker Y] supplied assembly documentation. Thank you very much. I value your efforts.''}, or \textit{``New error as of Dec. 2019. Appreciation to [Maker Z] for documenting this issue.''} Again, it is important to note that these attributions are done manually, relying entirely on the Thing author to uphold this best practice.

\section{Discussion}

This section discusses the main implications of our work, which are targeted towards the participants, platform designers, and researchers of OSH projects.

\subsection{Best Practices for Future Open Hardware Projects}

Among existing accounts of OSH, DrawBot stands out as a successful and sophisticated example, in terms of fostering a long-lived community and continuously-evolving design. As such, we believe there are several lessons to be learned from this case that can be applied to future OSH projects to improve collaborative design. In the following paragraphs, we discuss three unique facets of the DrawBot project that, in our view, played a crucial role in its success.

The first facet is the use of high-quality documentation, which has been highlighted as key to successful OSH projects~\cite{antoniou_defining_2022}. Despite the lack of formal documentation support features, the DrawBot author managed to create this documentation through iterative additions to the Thing description. Notably, the DrawBot's description was an order of magnitude longer than that for other Things, reflecting the depth and richness of available documentation. Given DrawBot's complexity, which involves designing, troubleshooting, and integrating both physical and digital artifacts, the comprehensive documentation outlining assembly instructions and addressing known challenges was exceptionally valuable. Strong documentation is key to retaining newcomers to OSS communities~\cite{fronchetti_contributing_2023,warncke-wang_increasing_2023}, and this may have contributed to the sustained engagement of newcomers in this project, contrasting with previous studies which found that makers tend not to continuously engage with a specific Thing~\cite{ludwig_designing_2018}.

The maintainer is significant in any open-source context, responsible for overseeing project development, reviewing contributions, and ensuring the overall sustainability of the project~\cite{gousios_work_2015}. The significance of the maintainer is particularly evident in the case of the DrawBot. Since there are no existing best practices for collaborative design on collective design platforms like Thingiverse, the maintainer -- or author in this case -- must assume a leadership role to establish norms, suggest tasks, and provide general guidance to the community. The DrawBot author conversed with every single DrawBot contributor (Figure~\ref{collaboration_network}), and this responsiveness likely fostered confidence in others about the quality and sustainability of the project. Importantly, responding to each newcomer is a laborious task~\cite{geiger_labor_2021}, and future work should explore workload management.

Alongside responsiveness, another factor potentially contributing to DrawBot's success was the author's interdisciplinary expertise spanning hardware, electronics, and software components. While this multidisciplinary expertise by the author was evident in their design contributions to the DrawBot, our study lacks the necessary data to confirm if a multidisciplinary project maintainer is the key to successful OSH collaboration. A promising area for future exploration is the impact of multidisciplinarity on OSH collaboration, which we discuss in Section~\ref{research_implications}. 

Although the DrawBot project succeeded with clusters of collaborators contributing to the same type of component, prior work suggests an alternative best practice. Salehi et al., in a study of software development, argue that rotating individuals within clusters of design tasks can be beneficial to collaboration, inviting more diverse perspectives and cross-pollination of ideas~\cite{salehi_hive_2018}. This difference in approach may stem from inherent distinctions between hardware and software projects, such as the nature of the modularity, which is closely tied to product architecture~\cite{gavras_mapping_2021}. 

Overall, DrawBot's success can be attributed to high-quality documentation, responsiveness, and interdisciplinarity. Future OSH maintainers could consider these best practices a valuable starting point when initiating similar collaborative design projects.



\subsection{Design Implications for OSH Platforms}
One of our study's contributions is identifying challenges that hindered the collaborative design process. Here, we see that even in successful OSH projects, tooling limitations hold project potential back. In an effort to minimize tooling-related challenges for future OSH projects, we discuss five design implications for the platform builders of OSH communities, aiming to improve awareness and traceability.

The first design implication is to make documentation more accessible and traceable. As presented in Section~\ref{scattered_collaboration}, contributors often struggle to maintain awareness of essential documentation scattered across various links, some of which lead to external sites such as GitHub and YouTube. While the maintainer tried to centralize documentation links in the Thing description, it is still cumbersome to explore content from disparate links. Similar challenges occur in OSS community sites, like Stack Overflow, and researchers have made recent contributions to increase the visibility of repeatedly-shared links~\cite{liu_exploratory_2021}, and automatically align documentation with relevant questions in discussion posts~\cite{pudari_aligning_2023}. OSH platforms may also benefit from more automated approaches for navigating documentation or summarizing documentation content. Although solutions from the software literature may assist in summarization methods for text-based content (e.g., assembly instructions), more advanced techniques are required to analyze images and CAD models~\cite{stemasov_shapefindar_2022}. 

Next, platform builders should address the poor navigation and traceability of the remixing graph, which makes it challenging for makers to find the necessary information. Nickerson relates this challenge to the \textit{theory of information foraging}~\cite{gero_collective_2015}, which aims to understand the way humans search, gather and consume information, often online~\cite{pirolli_information_1995}. Generally, the longer a user spends searching for a new idea, the less likely that it will be found~\cite{pirolli_information_1995}. Consequently, if new designs are difficult to find in remixing graphs, there is a risk that contributions can go unnoticed (Section~\ref{lost_contributions}). Thus, it is imperative to improve the design of the remixing graph to provide users with a better overview of remixes and sources. Drawing on the types of design-related contributions we identified, there could be value in organizing remixes by: (1) component, e.g., grouping all parts related to the pen holder; (2) product family, e.g., grouping all the files that pertain to a downsized DrawBot; and (3) version, e.g., grouping parts belonging to the same revision. 

Along the same lines, contributions can be buried in lengthy comment threads. A fruitful area for future work could involve developing methods to summarize the knowledge exchanges and outcomes of such conversations, inspired by OSS research~\cite{gilmer_summit_2023}. Such a summarization tool could automatically identify feature requests for the next design iteration. Furthermore, it would streamline the project onboarding process by automatically directing newcomers to relevant conversation threads, alleviating the burden on the maintainer to respond to each question individually. 

The next design implication motivates the development of an issue tracker to identify/flag and react to bugs/errors. In the DrawBot project, all of the tasks for tracking issues were done manually, which is tedious and error-prone. Designing a system similar to the GitHub issue tracker would streamline reporting, updating, and resolving issues~\cite{Bissyande2013}. 

Finally, OSH platforms need sophisticated version control and pull request functionalities, allowing makers to contribute directly to the Thing without resorting to workarounds, such as remixing and copying files from the remix into the original Thing as a pseudo-merge. There are many shortcomings with this method: the manual merge involves many steps; only the original author can edit the Thing; and there is no version history, further impairing contributors' awareness of the design evolution. Branching and merging for CAD files is indeed emerging as a promising tool for CAD collaboration, but there are still essential features lacking, such as selective merging~\cite{cheng_user_2023}.

Our work contributes to understanding awareness and traceability needs in OSH projects, providing specific areas for future development. \textcolor{black}{Implementing these design insights will not only aid in overcoming awareness challenges within OSH projects but may also be relevant for broader open collaboration contexts, such as software development~\cite{dabbish_social_2012} or collaborative writing~\cite{keegan_history_2013,geiger_work_2010}.}

\subsection{Practical Implications for OSH Researchers}\label{research_implications}

To our knowledge, this study is the first to analyze a network of connected designs on a collective design platform like Thingiverse, aiming to understand the challenges and opportunities of a collaborative design project. In this section, we aim to provide practical guidance for OSH researchers who may wish to conduct similar empirical investigations of collaboration processes. First, we present some heuristics that allowed us to recognize core contributors, followed by a discussion on navigating multidisciplinarity in OSH projects.

\subsubsection{Identifying Collaborators}
By qualitatively coding comment threads, we were able to gain deeper insight into the nature and context of design changes, which are not captured in remix activity alone. One notable finding is that the authors of remixes are not representative of all contributors. Since qualitative analysis can be time-consuming, making it impractical to read every comment for contributor identification, we have compiled a short list of heuristics that we believe can be valuable in future studies to more efficiently identify key contributors. Generally, we found that contributors were:
\begin{itemize}
    \item Makers who made the most comments.
    \item Makers who comment on two or more related Things, indicating that they are following the evolution of the design in remixes.
    \item Makers whose usernames were mentioned in the Thing description, often signifying acknowledgement of their contributions by the author.
    \item Makers who include a Thingiverse link in their comment. This usually indicated that they referred to a related design, which may not have been captured as a remix. 
\end{itemize}

Following and expanding on these heuristics can help researchers identify and analyze contributors to OSH projects more efficiently.

\subsubsection{Navigating Multidisciplinarity}

A product like DrawBot requires collaboration between makers with hardware expertise and makers with software expertise, but the platforms supporting them are distinct and disconnected. In the DrawBot project, we observed version tracking and evolution of the hardware components occurring on the Thing page, while software contributions were occasionally made on GitHub. It is possible that the development processes can diverge on different platforms. Therefore, OSH researchers conducting similar investigations should consider tracking design and collaboration activity on concurrent platforms, which might offer interesting insights into how multidisciplinarity affects OSH collaboration.

Below are some open questions regarding multidisciplinarity to tackle in future work: 
\begin{itemize}
    \item Are the designers collaborating on hardware platforms (e.g., Thingiverse) also involved in collaborations on software platforms (e.g., GitHub)?
    \item Do the platforms separate software-focused and hardware-focused designers? 
    \item How might collaboration practices differ, considering that GitHub is a more sophisticated version control system than Thingiverse, which is a collective design platform?
\end{itemize}

With these practical implications, our goal is to assist OSH researchers interested in conducting case studies in a similar fashion. Building on our work with additional case studies will contribute to a more nuanced understanding of OSH collaboration to the CSCW community.




\subsection{Limitations \& Future Work}

As with any case study, our goal is not to be generalizable to all OSH projects. Instead, we aimed to uncover insights potentially relevant to other projects through an in-depth examination of collaborative practices within an advanced OSH design project. 
Therefore, as was our intention, this project's characteristics are not representative of all the designs in collective design platforms. 

Next, while our project selection allows us to achieve our research objective, it only provides insights into one online community. Future work should compare our findings with other collaborative design projects, to gain a more comprehensive understanding of OSH collaboration.

Another limitation pertains to how we define a project and what constitutes a successful one. Given the novelty of our approach, we based our inclusion criteria (e.g., remixes and ancestors) and success metrics (e.g., number of likes, comments, remixes) on insights derived from related literature~\cite{oehlberg_patterns_2015,alcock_barriers_2016,ozkil_collective_2017,antoniou_defining_2022,papadimitriou_towards_2014,novak_500_2020,voigt_not_2018,baik_role_2022}. It is in the interest of future work to validate our methods and formalize project-level boundaries on platforms like Thingiverse.

Furthermore, our study is limited by the type of activity trace data~\cite{shi_value_2023} available through Thingiverse's API. As a result, we cannot analyze design processes that occurred outside of the Thingiverse platform, such as in-person assembly activities. In future work, we plan to conduct qualitative interviews with the collaborators of an OSH project, to gain these additional insights.

Finally, qualitative analysis can be vulnerable to the researcher's bias. To address this concern, we analyzed multiple data sources, including comments, descriptions, and actual design changes in CAD models, to triangulate our findings, which is recommended for case study research~\cite{yin_case_2018}. 

\section{Conclusion}
In this work, we conducted a detailed case study of collaborative design practices in the DrawBot project on Thingiverse. Using our mixed-methods approach, we analyzed discussions in comment threads, design contributions in remixing relationships, and changes made to various documentation, to understand how collaboration occurred in a successful and complex OSH project. 

Our research revealed that designers collaborated on various tasks, including designing new features, addressing errors, enhancing performance, and replacing components of the DrawBot. Collaborative efforts ranged from short-term discussions involving a few makers to address noise reduction to more extensive collaborations lasting 11 months, focusing on resolving significant software errors. Practical considerations -- those related to designing a physical product, such as material availability, costs, and the use of standardized electronic components -- played a crucial role in shaping these collaborations. In addition, not all collaborations resulted in a change to the DrawBot design; many discussions centred around process-related contributions, such as documentation improvements and troubleshooting assistance. Lastly, our study highlighted unique approaches exercised by the DrawBot community to overcome limitations within the Thingiverse platform, such as tracking version history. 

This case study sheds light on the resilience and resourcefulness of the DrawBot community, demonstrating that a collaborative OSH design project can indeed thrive, despite inadequate support from existing platforms. However, in order to advance the scope of OSH, our study also identifies various areas for improvement. We contribute best practices for OSH project maintainers to enhance collaboration efficiency, design implications for platform builders to address current challenges, and practical implications for guiding researchers in expanding OSH research within CSCW. With each step taken towards improving design collaboration, we push the boundaries of what OSH innovation can achieve.

    


\bibliographystyle{ACM-Reference-Format}
   \bibliography{sample-base}

\appendix

\section{Components of the DrawBot}\label{appendix_codebook}
Table~\ref{codebook} summarizes and describes the components of the DrawBot.

\begin{table}[h]
\footnotesize
  \caption{Components required to create and use the Drawbot, including hardware, electronics, and software.}
  \begin{center}
  \label{codebook}
  \begin{tabular}{p{.5in} p{.7in}>{\raggedright\arraybackslash} p{1.85in} p{1.85in} }
      \toprule
    Category&Sub-category&Description&Example(s) from Dataset\\
    \midrule
    Hardware&Pen Holder&A gripping device that securely holds the drawing tool.&\textit{``I want to explore swivel blade mount to cut materials.''}\\ 
    &Pen Lift&A mechanism that raises and lowers the drawing tool in the Z-direction.&\textit{``Your pen isn't lifting enough. Try increasing the angle to servo from 35 to 45 or 60.''}\\ 
    &Guide Rails&Linear rods that provide stability and directional guidance for the clamshell.&\textit{``Are the rods prone to bending when the Y-axis is fully extended?''}\\ 
    &Clamshell&A component that moves along the guide rails to enable precise X- and Y-axis movement of the pen holder.&\textit{``I've trimmed some material from the upper clamshell that was interfering with the lower clamshell posts.''}\\ 
    &Stepper Motor Mount&A housing for the stepper motors driving the movement of the clamshell.&\textit{``Does putting the motors on top lead to more vibrations compared to on the bottom?''}\\ 
    &Belt Guide&A belt and pulley system that transfers motion from the stepper motors to the clamshell.&\textit{``Any suggestions on how to increase belt tension? Maybe by adding a belt tensioner attachment?''}\\ 
    \midrule
    Electronics&Arduino Uno&A microcontroller board that interprets and translates drawing instructions for the CNC Shield.&\textit{``A standard Uno lacks the horsepower to read files from an SD card and plot them. You would need a Mega Arduino.''}\\ 
    &CNC Shield&An extension board mounted on the Uno that precisely controls the motors.&\textit{``Make sure the stepper drivers are correctly placed on the CNC board.''}\\ 
    \midrule
    Software&Inkscape \& MI Extension&A vector graphics software used to create and edit vector graphics; extensions export graphics as G-code.&\textit{``I'm having problems exporting intricate designs (like more than 5 paths); Inkscape crashes or is unresponsive.''} \\ 
    &UGS (Universal G-Code Sender)&A program that communicates to the Uno to send G-code, enable manual control, and reset the DrawBot.& \textit{``Ensure that when using UGS to move the pen in either the X or Y direction, it shouldn't follow a diagonal path.''} \\ [10pt]
    &GRBL (G-code Real-time Boot Loader)&A firmware program that runs on the Arduino Uno that interprets G-code to control the DrawBot's movements.&\textit{``You need to use the special version of GRBL that can manage commands for raising and lowering the pen.'' }\\ 
    &Other Software&General software needed to design, develop, or run the DrawBot.&\textit{``Check and make sure you're running Python version 3.X on your computer.''}\\ 
  \midrule
\end{tabular}
\end{center}
\end{table}

\end{document}